\documentclass[preprint]{aastex}
\usepackage{graphicx}
\usepackage{lscape}

\shorttitle{}
\shortauthors{}
\begin{document}

\title{Apparent Faster-Than-Light Pulse Propagation in Interstellar Space: A new probe of the Interstellar Medium}

\author{F. A. Jenet\altaffilmark{1}, D. Fleckenstein\altaffilmark{1}, A. Ford \altaffilmark{1}, A. Garcia\altaffilmark{1},R. Miller\altaffilmark{1}, J. Rivera\altaffilmark{1}, K. Stovall\altaffilmark{1}}

\altaffiltext{1}{Center for Gravitational Wave Astronomy,
University of Texas at Brownsville, TX 78520 (merlyn@phys.utb.edu)}

%%%%%%%%%%%%%%%%%%%%%%%%%%%%%%%%%%%%%%%%%%%%%%%%%%%%%%%%%%%%%%%%%%%%%%%%%%%%%

\begin{abstract}

Radio pulsars emit regular bursts of radio radiation that propagate
through the interstellar medium (ISM), the tenuous gas and plasma
between the stars. Previously known dispersive properties of the ISM
cause low frequency pulses to be delayed in time with respect to high
frequency ones. This effect can be explained by the presence of free
electrons in the medium. The ISM also contains neutral hydrogen which
has a well known resonance at 1420.4 MHz. Electro-magnetic theory
predicts that at such a resonance, the induced dispersive effects will
be drastically different from those of the free electrons. Pulses
traveling through a cloud of neutral hydrogen should undergo
``anomalous dispersion,'' which causes the group velocity of the
medium to be larger than the speed of light in vacuum. This
superluminal group velocity causes pulses containing frequencies near
the resonance to arrive earlier in time with respect to other
pulses. Hence, these pulses appear to travel faster than light. This
phenomenon is caused by an interplay between the time scales present
in the pulse and the time scales present in the medium. Although
counter-intuitive, it does not violate the laws of special
relativity. Here, we present Arecibo observations of the radio pulsar
PSR B1937+21 that show clear evidence of anomalous dispersion. Though
this effect is known in laboratory physics, this is the first time it
has been directly observed in an astrophysical context, and it has the
potential to be a useful tool for studying the properties of neutral
hydrogen in the Galaxy.

\end{abstract}


%%%%%%%%%%%%%%%%%%%%%%%%%%%%%%%%%%%%%%%%%%%%%%%%%%%%%%%%%%%%%%%%%%%%%%%
\section{Introduction}

A radio pulsar is a rapidly spinning neutron star that emits a beam of
radio radiation. This highly columnated beam of radiation is rotated
into and out of the line-of-sight of a distance observer as the star
itself rotates about a fixed axis. Hence, an observer sees a periodic
train of pulses arriving at regular intervals of time. The nature of
these pulses gives astronomers a unique tool to study various physical
phenomena including electro-magnetic propagation effects in the
interstellar medium (ISM).

Classically, pulsar pulses are affected by four basic ISM related
propagation effects which can alter a pulse's observed properties:
dispersion, absorption, Faraday rotation, and scattering
\citep{mt77}. Dispersion is caused by free electrons in the ISM
\citep{cwf+91}. Pulse absorption is caused by clouds of neutral
hydrogen which absorb energy near the spin-flip resonance frequency of
1420.4 MHz \citep{wsx+08, jkw+01}. The presence of a magnetic field
together with free electrons cause Faraday rotation, which changes the
orientation of the electric field's polarization vector
\citep{hml+06}. Scattering effects are created by free electron
density inhomogeneities \citep{ars95}. Recently, \citet{wjk+05} found
evidence for stimulated emission driven by pulsar
emission. Interstellar hydroxl (OH) clouds amplify the intensity of
pulses with energy near the OH resonance. These four phenomena,
together with the modern discovery of pulsar driven stimulated
emission, encompass all previously known and studied ISM propagation
effects.

This paper reports on the discovery of a new ISM propagation
effect. The arrival times of pulses are seen to be delayed or advanced
over and above the standard free-electron dispersion delay near the
spin-flip resonance frequency of neutral hydrogen. In the next
section, the physics of pulse dispersion is reviewed and the expected
pulse time delay is calculated for the case of a cloud of neutral
hydrogen with a thermal (i.e. Gaussian) velocity distribution. The observations and
data analysis techniques used to measure this effected are presented
in \S 3. The results and conclusions are given in \S 4.

\section{Pulse dispersion in the ISM}

The free electrons in the ISM induce a frequency dependent index of
refraction, the ratio of the vacuum light propagation speed to the
actual phase velocity in the medium. The index of refraction may be
calculated directly from the ``dispersion relationship'', the equation
which relates the wave number, $k$, to the wave's frequency, $f$. For
the case of free electrons in the ISM, the dispersion relationship takes the
following form assuming cgs units:
\begin{equation}
k  = \frac{2 \pi f}{c}\sqrt{1 - \frac{n_e e^2}{ \pi m_e f^2}},
\label{disp_cp}
\end{equation}
where $n_e$ is the electron number density, $e$ is the electron
charge, and $m_e$ is the electron mass. The index of refraction
is given by $k c/ 2 \pi f$, where c is the speed of light in
vacuum. From the above dispersion relationship, one can calculate the
index of refraction for free electrons:
\begin{equation}
n(f) = \sqrt{1 - \frac{n_e e^2}{ \pi m_e f^2}}.
\label{index_cp}
\end{equation}
For frequencies greater than $\sqrt{n_e e^2/ \pi m_e}$, the so called
``plasma frequency'', the index of refraction is less than unity. This
implies that the phase velocity is greater than the speed of
light. This is not a violation of any physical principal since it is
well known that information propagates at the ``group velocity,''
which is defined as the derivative of the angular frequency, $2 \pi
f$, with respect to the wavenumber, $k$. From equations \ref{disp_cp}
and \ref{index_cp}, it can be shown that 
\begin{equation}
v_{\mbox{group}} = 2 \pi df/dk = c n(f).
\end{equation}
Since $n(f)$ is less than unity for frequencies above the plasma
frequency, the group velocity is less than the speed of light in
vacuum. The functional form of $n(f)$ tells us that high frequency
pulses travel faster than low frequency pulses. Using the group
velocity, one can calculate the time, $\Delta$, it takes for a pulse
to travel from the pulsar to the telescope. Assuming that the free
electron density is sufficiently low ($n_e << 10^7$ cm$^{-3}$
(f/1GHz)$^2$), which is true for observing frequencies above 22kHz,
one can approximate $\Delta$ by its Taylor series expansion out to
first order in $n_e$. One then finds that the propagation time may be written as a sum of two delays:
\begin{equation}
\Delta = \Delta_v + \Delta_{fe},
\end{equation}
 where 
\begin{eqnarray}
\Delta_v &=& D/c, \label{t1}\\
\Delta_{fe}(f) &=& \frac{e^2 n_e D}{2 \pi m_e c f^2}. \label{t2}
\end{eqnarray}
$D$ is the distance between the Earth and the pulsar, $\Delta_v$ is the
vacuum propagation time, and $\Delta_{fe}(f)$ is the added frequency
dependent delay due to the free electrons. Note that in order to
simplify the discussion, it is assumed that the free electron
density is uniform. The product $n_e D$ is known as the dispersion
measure. If the free electron density varies along the line of sight,
this product is replaced by the integral $\int_0^D n_e(l) dl$.

The presence of bound electrons can alter the index of refraction of a
medium significantly near the resonance frequency
\citep{jac75,som54}. $n(f)$ will have an imaginary part which is
responsible for the absorption as well as a real part which determines
the propagation properties. At resonance, the group velocity becomes
greater than the speed of light in vacuum. This effect, a result of a
phenomenon known as anomalous dispersion, can cause pulses to appear
to travel faster-than-light without violating causality
\citep{wkd00}. Although the effects of anomalous dispersion have been
measured in ground based laboratories \citep{slb+06,dzw01}, they have
not been observed in an astrophysical context. Pulsar observations
offer a unique opportunity to directly measure the effects of anomalous
dispersion using the electron spin-flip transition in neutral
hydrogen.

Consider the case of a medium made up of a cloud of neutral
hydrogen only. First, lets assume that the atoms are not moving with
respect to the observer. In this case, the dispersion relationship
takes the form \citep{jac75}:
\begin{equation}
k^2c^2 = (2 \pi f)^2 - \frac{\left(n_h e^2 s_h/\pi m_e\right)(2 \pi f)^2}{f^2 - f_0^2 + i 2\frac{f}{\tau}}
\label{disp2}
\end{equation}
where $n_h$ is the number density of hydrogen, $s_h$ is the
``oscillator strength,'' $\tau$ is the life time of the spin-flip
transition, and $f_0$ is the resonant frequency of the
transition. Note that $n_h$ is the number density only when the
spin temperature is zero. Otherwise, its the difference between the
number density of atoms with its electron spin in its ground state
minus the number density of atoms with its electron spin in its
excited state. 
%The
%peak amplitude of the resonant term is given by $(n_h e^2 s_h/2 \pi
%m_e)f_0 \tau$. As long as the density, $n_h$, is less than ??, as
%is typical for the ISM, the magnitude of the second term in equation
%\ref{disp2} is always much less than unity. 
From the above dispersion relationship, the squared index of refraction is
given by
%may be accurately represented by its Taylor series expansion out to first order in $n_h$:
\begin{equation}
n(f)^2 = 1 - \frac{\left(n_h e^2 s_h/\pi m_e\right)}{f^2 - f_0^2 + i 2\frac{f}{\tau}}
\label{index2}
\end{equation}
It will be assumed that we are only interested in frequencies
near $f_0$. Since $f_0 = 1.4$ GHz and the
observing bandwidth is of order 1 MHz, $|f - f_0|/f_0 << 1$.  
%Also,
%we will make use of the fact that $1/\tau$ is of order $10^{-15}$ Hz
%and therefore much less than $|f - f_0|$.  
In this case, one can approximate $f^2 - f_0^2$ as $2 f (f - f_0)
$. With this approximation, the squared index of refraction becomes:
\begin{equation}
n(f)^2 = 1 - \frac{\left(n_h e^2 s_h/2 \pi m_e f\right)}{f - f_0 + i \frac{1}{\tau}}
\end{equation}
If the atoms were moving with respect to the observer at speeds small
compared to the speed of light, $f_0$ in the denominator of the above
expression would be replaced by its Doppler-shifted value of $f_0(1-
v/c)$ where $v$ is the radial velocity. If one were to add in another
set of atoms moving at a velocity different from the first set, the
index of refraction would be adjusted simply by adding another term of
a form identical to the second term in the above expression except
with different values of $n_h$ and $f_0$. Following this logic, the
squared index of refraction for the case of a cloud of neutral hydrogen with a
thermally distributed set of radial velocities takes the form:
\begin{equation}
n(f)^2 = 1 - \frac{n_h e^2 s_h}{2 \pi m_e f} \frac{1}{\sqrt{2 \pi} f_d}\int \frac{ e^{\frac{-( f' - f_c)^2}{2 f_d^2}}}{f - f' + i \frac{1}{\tau}} df',
\label{index2_approx}
\end{equation}
where $f_c = f_0(1 - v_c/c)$, $v_c$ is the average velocity of the
cloud, $f_d = f_0\sqrt{ k_b T/m_h c^2}$ is the thermal frequency width
of the cloud, $k_b$ is Boltzmann's constant and $T$ is the kinetic
temperature of the cloud. 

The integral on the right hand side of equation
\ref{index2_approx} may be evaluated analytically in terms of the
$w(z)$ function defined in \citet{as70}, section 7. In terms of
$w(z)$, the squared index of refraction becomes:
\begin{equation}
n(f)^2 = 1 + \frac{i n_h e^2 s_h}{2\sqrt{2\pi} m_e f f_d}w\left(\frac{f-f_c + i/\tau}{\sqrt{2}f_d}\right)
\label{index2_w}
%n(f) = 1 - \frac{n_e e^2}{ 2 \pi m_e f^2} + \frac{i \sigma_0}{2 \pi c f}w\left(\frac{f-f_0}{f_d}\right),
\end{equation}
%where $f_0$ is the rest frequency of the spin-flip resonance,
%$\sigma_0$ is the peak absorption coefficient, $f_d = f_0 \sqrt{2 k
%  T/m_h c^2}$ is the frequency width of the neutral hydrogen line
%profile, T is the cloud temperature, and $m_h$ is the mass of
%hydrogen. 
%where the function $w(z)$ in the last term is given by (See
%\citet{as70}) $w(z) = e^{-z^2}[1-\mbox{erf}(-iz)]$ where erf($z$) is
%the complex error function. 
Since $w(z) \leq 1$ provided that the imaginary part of $z$ is greater
than zero, the maximum amplitude of the second term in equation
\ref{index2_w} is given by $n_h e^2 s_h/2 \sqrt{2\pi} m_e f_d^2$. A
full quantum mechanical treatment shows that $s_h = hf_0/2 m c^2$
where h is Planck's constant \citep{cs63}. Hence, the maximum amplitude
is much less than unity as long as $n_h << 10^{21}
\mbox{cm}^{-3}$. Under this assumption, we can accurately approximate
$n(f)$ by:
\begin{equation}
n(f) = 1 + \frac{i n_h e^2 s_h}{4\sqrt{2\pi} m_e f f_d}w\left(\frac{f-f_c + i/\tau}{\sqrt{2}f_d}\right)
\end{equation}

In order to simplify further the above expression, we will ignore the natural
line width term, $i/\tau$, since it is much less than the Doppler
broadening width $f_d$. Also, we will define $\sigma_o= \sqrt{\pi}
n_h e^2 s_h/\sqrt{2} m_e f_d c$. The index of refraction may
now be written as:
\begin{equation}
n(f) = 1 + \frac{i \sigma_0 c}{4 \pi f} w\left(\frac{f-f_c}{\sqrt{2}f_d}\right)
\label{index2_w_simp}
\end{equation}

There were several approximations used to derived the above expression
for the index of refraction. These approximations are summarized
here. First, it is necessary that $1/\tau << f_d << f_0$. Since $f_d$
is determined by the cloud temperature, it can be shown that the above
expression for $n(f)$ is valid as long as $ 10^{-26} \mbox{K} << T <<
10^{14}\mbox{K}$. Second, the number density of hydrogen atoms in the
cloud must be less than $10^{21} \mbox{cm}^{-3}$. Both of these
conditions are easily satisfied in the ISM where the cloud
temperatures range from 10 K up to $10^4$~K and the densities range from
less than $0.01 \mbox{cm} ^{-3}$ to of order $100 \mbox{cm}^{-3}$
\citep{bg02}. Note, there is evidence of HI clouds with densities up
to $10^4 \mbox{cm}^3$ \citep{jkw+03}, but even this relatively high density is still
well within the regime of validity for equation \ref{index2_w_simp}.

The index of refraction determines both the rate at which the
intensity of the wave-packet decreases as well as the time it takes for
the wave-packet to travel through the medium. The intensity of the wave-packet
decreases as $e^{-\sigma(f)D}$ where $\sigma(f)$ is the frequency
dependant absorption coefficient and $D$ is the distance traveled
through the cloud. The absorption coefficient is determined from the
imaginary part of the index of refraction: $\sigma(f) = \mbox{Im}(4
\pi f n(f)/c)$. From equation \ref{index2_w_simp}, one finds that
$\sigma(f)$ is proportional to the real part of $w(z)$, which is
identically equal to $e^{-z^2}$ when $z$ is purely real (See
\citet{as70}, section 7.1). Given $\sigma(f)$ and the length scale of
the cloud, $D_c$, the optical depth of the cloud, $\tau$, may be
calculated:
\begin{equation}
\tau(f) = \sigma(f)D_c = \tau_0 e^{-\frac{(f - f_c)^2}{f_d^2}},
\end{equation}
where $\tau_0 = \sigma_0 D_c$.  

The real part of $n(f)$ determines the propagation properties of the
wave. The propagation time across a cloud is determined by
$\Delta(f) = D_c/v_{\mbox{group}}(f) = (D_c/c) \mbox{Re}[dk/df]/2 \pi = (D_c/c) d
(f n(f))/df $. For the index of refraction given in equation
\ref{index2_w_simp}, $\Delta(f)$ may be written as
\begin{equation}
\Delta(f) = \frac{D_c}{c} + \frac{\sigma_0 D_c}{4 \sqrt{2}\pi f_d} \mbox{Re}\left[ i w'\left(\frac{f-f_c}{\sqrt{2}f_d}\right)\right]
\end{equation}
where $w'(z) = d w(z)/dz$. As for the case of the free electrons, one
can see that the propagation time may be written as a sum of the
vacuum delay term and a term involving the hydrogen resonance:
$\Delta(f) = D_c/c + \Delta_{rd}(f)$. \citet{as70} gives an expression
for $w'(z)$ in terms of $w(z)$: $w'(z) = - 2 z w(z) + 2
i/\sqrt{\pi}$. Using this, the resonant dispersion component of the
time delay, $\Delta_{rd}(f)$, may be written as:
%\begin{equation}
%\Delta(f) = \frac{D_c}{c} 
%+ \tau_0 \left(\frac{f - f_c}{4 \pi f_d^2}\right) \mbox{Im}\left[ w \left(\frac{f-f_c}{\sqrt{2}f_d}\right)\right]
%- \frac{\tau_0}{2 \sqrt{2}\pi^{3/2} f_d} 
%\end{equation}
 \begin{equation}
\Delta_{rd}(f) = \tau_0 \left(\frac{f-f_c}{4 \pi f_d^2}\right) \mbox{Im}\left[w\left(\frac{f - f_c}{f_d}\right)\right] - \frac{\tau_0}{2 \sqrt{2}\pi^{3/2} f_d}.
\end{equation}

\begin{figure}
\epsscale{1.0}
\plotone{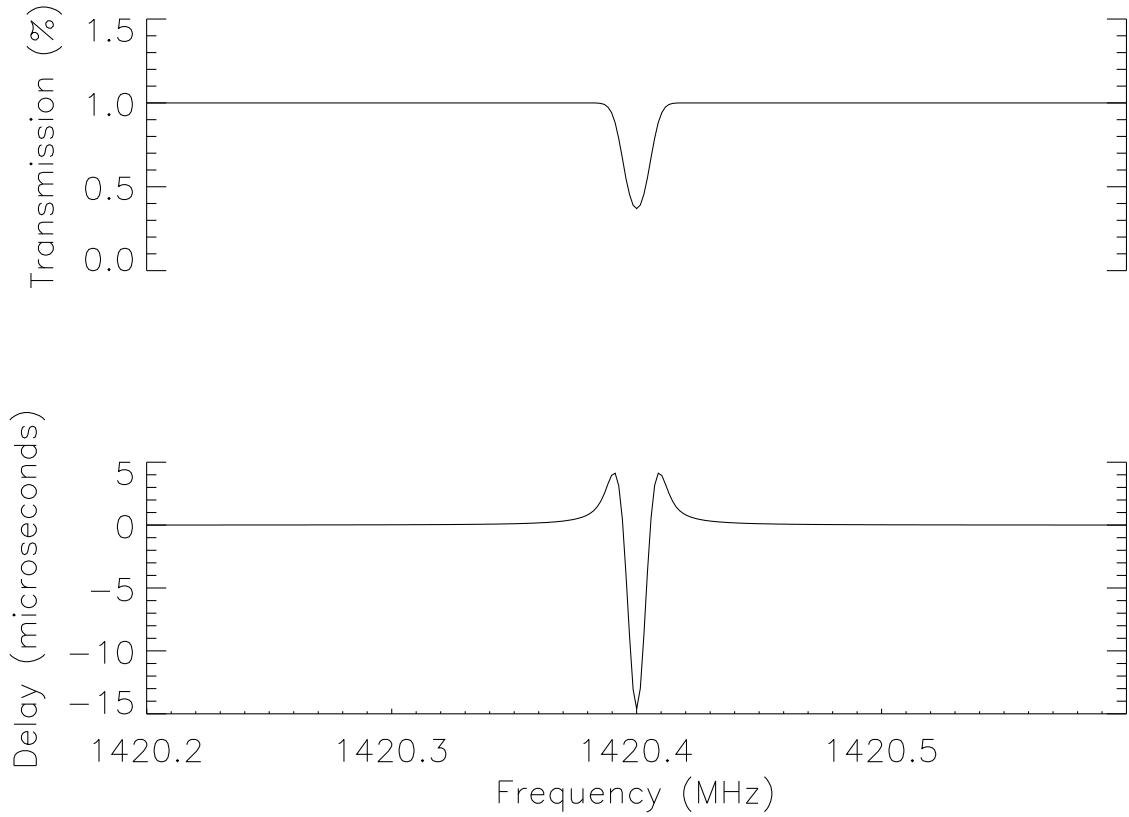}
\caption{\label{fig1}The expected absorption (top) and delay (bottom)
  spectrum caused by a cloud of neutral hydrogen (HI) in the ISM. The
  kinetic temperature is 100K and the optical depth is unity. At the
  spin-flip transition frequency, the delay is seen to be negative,
  corresponding to a pulse advance. Pulses at this frequency appear to
  arrive earlier then pulses at frequencies off resonance. Note, the
  free electron dispersive effects are not included in the above delay
  spectrum.}
\end{figure}

As an example of what we might expect see in actual data, Figure
\ref{fig1} shows the absorption spectrum given by $e^{-\tau(f)}$ and
the frequency dependent delay, $\Delta_{rd}(f)$, for a single hydrogen
cloud. The peak optical depth is unity and the temperature is
$100K$. Near the peak of the resonance, $\Delta_{rd}(f)$ is less than
zero, implying that the pulses arrive earlier in time. 

The unique shape of the frequency dependent delay warrants further
discussion. The time delay is proportional to $1/v_{group} = n(f) + f
dn(f)/df$. For most frequencies, the dispersion is 'normal' where
normal dispersion is defined as $dn(f)/df >0$. Within a small range of
frequencies near resonance, $dn(f)/df <0$ and the dispersion is said
to be 'anomalous'. If the magnitude of $dn(f)/df$ is sufficiently large, as is the case
for HI resonant dispersion, the delay will become negative. Since the
dispersion is normal for most frequencies and since $dn(f)/df$ is
continuous, there must be regions on either side of the anomalous
dispersion region where the dispersion transitions back to normal. It
just so happens that the transition regions are still close enough to
resonance so that normal dispersion is enhanced
thus producing the delay 'wings' to either side of the
anomalous dispersion negative delay peak. Further discussion of this
phenomenon may be found in \citet{jac75}, section 7.8.

%At this point, one could side-step the issues of causality and
%relativity by noting that the advanced pulses are arriving earlier
%than pulses that are already delayed by the free-electon component of
%the ISM. Since the pulse advance is much smaller than the delay
%induced by the free electrons, the pulses are not traveling faster
%than light in vacuum. Unfortunately, the fact of the matter remains
%that if the pulses are traveling through a region of space that is
%occupied predominently by neutral hydrogen alone, which is possible,
%then the pulses will be traveling faster than light. 

The reason that this pulse advance is not a violation of causality and
relativity is a subtle one. Put simply, the peak of the pulse does not
necessairly carry information. When the rising edge of the pulse
enters the plasma, it causes the atoms to start emitting radiation at
the same frequencies as that present in the wave-packet, but each with
a slightly different phase. This radiation interferes with the
radiation in the wave-packet in just the right way as to cause the
intensity to start to decrease earlier than it would have in
vacuum. Hence, the peak has advanced. It will not advance earlier than
the time the leading edge reached the plasma. 

Lets assume that we wanted to create a signal transmission device that
could send information faster than the speed of light using the
anomalous dispersion effect. We could imagine having a pulse generator
that, when a button is pressed, emits a wave packet with frequency
content near the HI resonance into an HI cloud. The pulse would travel
through the cloud and hit a receiver at some distance $D$ away. For
argument sake, lets assume that the pulse envelope is described by a
Gaussian function with a given width. The device is setup so that,
initially, it is not transmitting. As soon as the button is pressed,
the device starts to transmit the signal. If we let $E(x,t)$ represent
the electric field of the signal and say that the button is pressed at
$t=0$, then $E(0,t) =0$ for all times less than zero and it is
non-zero at $t>0$. It can be shown formally that $E(D,t)$ must be
equal to zero for all times $t<D/c$ \citep{jac75}. The transmission
device was programed to emit a pulse whose peak power occurs at some
known duration after the button is pressed. No matter what happens to
the peak of the pulse, it cannot arrive earlier than $D/c$. The peak
will have occured at some positive time after the button is
pressed. As it travels through the HI cloud, the peak may be moving
faster than light for some distance, but it will still arrive after a
time $D/c$, the earliest time when information about the button being
pressed could be received. It may be the case that the receiver
detects the signal only after a threshold intensity is reached. In
which case, the signal traveling through the HI cloud may be detected
before the signal traveling through vacuum, but the detection will
occur after a time $D/c$. Note that if the peak continued to move at a
rate fater than light, the peak would eventually arrive earlier than
$D/c$. In this extreme case, the assumptions used to calculate the
group velocity break down and the group velocity no longer describes
the propagation of the peak. As long at the time shift is small
compared to the width of the pulse, the group velocity will describe
the propagation of the peak of the pulse and the pulse shape itself
will not change \citep{gm70}.

In the ISM, there can be free electrons present together with
neutral hydrogen. In this case, the total time delay is a sum of the
free electron delay together with the time delays associated with each HI
cloud present along the line-of-sight to the pulsar. It should be
emphasized that the anomalous dispersion effects are completely
separate from the normal dispersion effects of the free electrons. One
may be tempted to side-step the whole causality issue discussed above
by saying that the pulses are advanced by a small amount with respect
to a much larger pulse delay introduced by the free electrons. If the
free electrons and the neutral hydrogen were somehow always forced to
occupy the same regions of space, this may be a valid argument. The
fact of the matter is that the neutral hydrogen may very well be
located in regions of space distinct from the free electrons. Pulses
traveling through the HI regions will appear to travel faster than
light. 

%For the case of
%typical free electron and hydrogen densities, the propagation time
%can again be written as a sum of independent terms:
%\begin{equation}
%\Delta(f) = \Delta_v + \Delta_{fe}(f) + \Delta_{ad}(f)
%\end{equation}
%where the first two terms are given by equations \ref{t1} and \ref{t2}
%and $tau_{ad}(f)$ is the added delay (or advance) due to the hydrogen
%spin-flip resonance. This added delay is given by
%\begin{equation}
%  \Delta_{ad}(f) = ??.
%\end{equation}
%Since $\Delta_{ad}(f)$ can be negative, pulses can appear to arrive
%earlier than they would under the influence of the free electrons
%alone. Figure \ref{fig1}, bottom panel, shows the expected anomalous
%dispersion delay, $\Delta_{ad}(f)$, as a function of frequency for
%a single neutral hydrogen cloud with a temperature of $100K$ and an
%optical depth on unity. A negative delay corresponds to a pulse
%advance. Together with this is the expected absorption spectrum. For
%typical ISM parameters, we expect pulse advances of order 10
%micro-seconds.

\section{Observations and Analysis}

Previous pulsar pulse arrival time observations have all been
consistent with free electron dispersion. Observations presented here
show the presence of anomalous dispersion at the resonance frequency
of the hydrogen spin-flip transition. Using the Arecibo\footnote{The
  Arecibo Observatory is operated by Cornell University under contract
  from the National Science Foundation}  305m radio telescope, pulsar
PSR B1937+21 was observed using the L-Band wide receiver centered on
1420.4 MHz. The data were recorded using the Wide-band Arecibo Pulsar
Processor (WAPP) using 128 frequency channels across a 1.5 MHz
bandwidth. Both polarizations were summed together. Data were taken
over three days. Observations lasted for 2 hours during the first
observing session, and 1.5 hours on the other days. For each day,
average pulse profiles were obtained in each channel using this
source's known phase parameters (i.e. its ephemeris). From the folded
profiles in each frequency channel, a pulse arrival time was obtained
by convolving the measured profile with an analytic profile consisting
of two Gaussian pulses. The parameters of the analytic profile were
obtained by fitting the measured pulsar profile. The effects of
dispersion were removed using standard incoherent dedispersion
techniques \citep{lk04}. Hence, the analytic profile was a good
representation of the average profile but without the added receiver
noise. See \citet{lk04} for details of pulsar timing techniques.

\begin{figure}
\epsscale{1.0}
\plotone{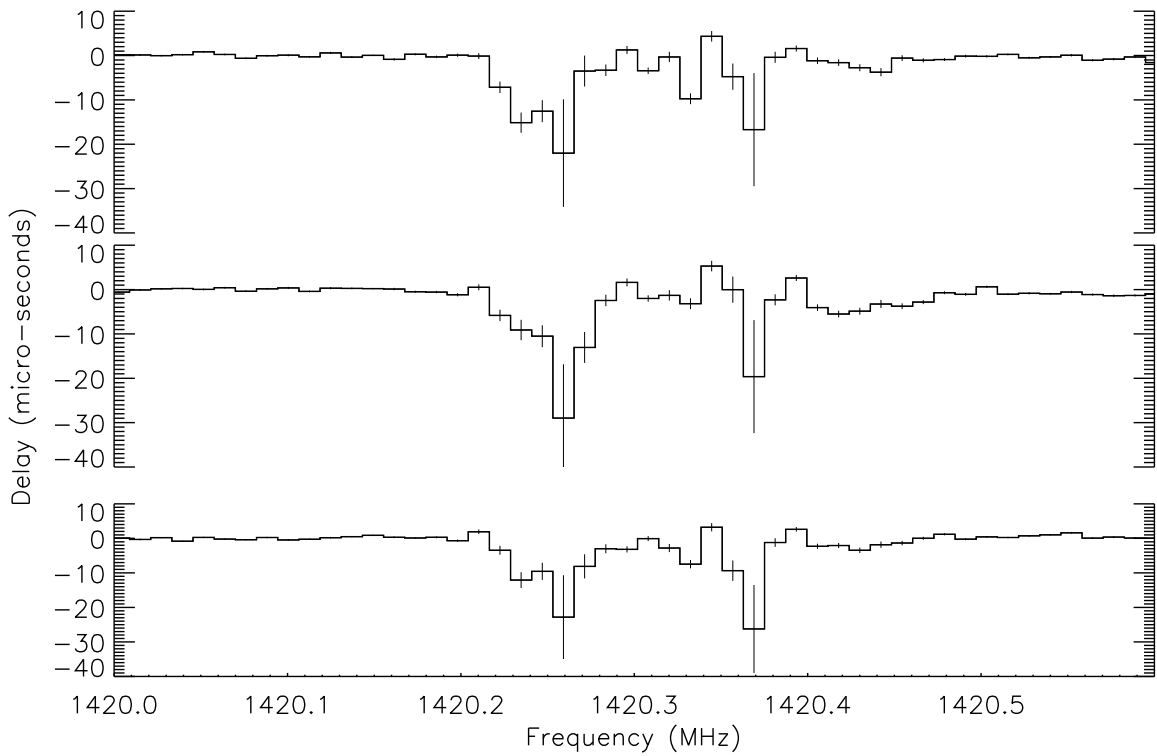}
\caption{\label{fig2}The delay spectra measured on three consecutive days. The top panel, day 1, was measured using a two hour observation. The remaining two days correspond to 1.5 hour observations. }
\end{figure}

Figure \ref{fig2} shows the measured pulse times-of-arrival relative
to the lowest frequency channel as a function of frequency. The three
panels correspond to the three different observing days. Over the 1.5
MHz band, the free electron dispersion delay, $\Delta_{df}$, can be
well approximated by a linear function of the frequency offset from
the center of the band. This linear trend was subtracted from the
delays presented in figure \ref{fig2}. The expected pulse advances are
clearly seen in all three days. The reproducible features are caused
by the structure and velocity of the hydrogen clouds located between
Earth and the pulsar.

Figure \ref{fig3} shows the measured anomalous dispersion delays
averaged over all three observing runs together with the hydrogen
emission and absorption spectra measured from the same data. Note that
these emission and absorption spectra are consistent with previously
published results by \citet{hks+83}. From the figure, it can been seen
that the structure in the anomalous dispersion ``delay spectrum'' is
consistent with the features seen in absorption. Note that
digitization effects are known to alter the absorption spectrum
\citep{wrb80}. Such systematic effects have been removed from the
absorption spectrum. 

It is possible that digitization effects could be affecting the delay
spectrum. In order to determine if digitization could be causing or
altering the measured delay spectrum, Monte-Carlo style simulations
were performed. Several million simulated pulses were generated by
modulating Gaussian noise with a Gaussian pulse profile. The effects
of the free electron and neutral hydrogen dispersion were added to the
signal by Fourier transforming the data into the frequency domain
using the Fast Fourier Transform technique (FFT), multiplying the
transformed data by a filter function derived from the appropriate
dispersion relationship, and then transforming the data back into the
time domain. The pulsar had a simulated period and dispersion measure
equal to that of PSR B1937+21. The simulated signal had a 1.5 MHz
bandwidth. The HI cloud was given a temperature of 100 K and a peak
optical depth of 1.6. Random Gaussian deviates were then added to the
data to simulate the effects of receiver noise. The noise level was
chosen to be consistent with the L-band wide receiver system. These data
were next digitized into n-bits and then broken up into 128 independent
frequency channels using an FFT based technique. The signal in each
channel was then squared and folded at the pulsar period. This
resulted in a folded filterbank in the same form as the actual data
taken with the WAPPs. Delay spectrum plots were then made in exactly
the same way as done using the real data. A delay spectrum was made
using 2-bits and compared to one made with 32-bits. No significant
differences were seen in the data sets. As an extreme test, the real
part of the HI cloud's index of refraction was set to zero, thus
removing the anomalous dispersion pulse arrival time
delays/advances. This test would determine if the digitization process
was somehow artificially causing systematic arrival time
delays/advances because of the frequency dependant absorption. No
significant features were seen in this delay spectrum. The above tests
were also done with no receiver noise added to the data. Again, no
significant features were seen in the delay spectrum. Hence, it was
concluded that digitization does not introduce significant artifacts
into the delay spectrum.

\section{Results and Conclusions}

The delay spectrum offers a completely new way to probe neutral
hydrogen clouds. The information obtained can be used together with
the absorption information to create a more accurate understanding of
a cloud's properties. It should be noted that the delay spectrum
carries different information than the absorption spectrum. The peak
of the absorption spectrum scales as the absorption parameter,
$\tau_0$, which is proportional to the HI column density divided by
$f_d$ (See equation \ref{index2_w_simp} and the definition of
$\sigma_0$ above it. Remember that $\tau_0 = \sigma_0 D_c$ and the
column density is given by $n_h D_c$). Since $f_d$ is proportional
to $T^{1/2}$, $\tau_0$ scales as $T^{-1/2}$ times the HI column
density. The peak of the delay spectrum, given by
\begin{equation}
\Delta_{peak} = -\frac{\tau_0}{(2 \pi)^{3/2} f_d},
\label{delay_peak}
\end{equation}
scales as $\tau_0/f_d$. Hence, the peak delay scales as the HI column
density times $T^{-1}$. This difference in the temperature dependence
will break the degeneracy between the kinetic temperature and the
column density that exists with only the absorption information
alone. The data presented here illustrates this idea.

\begin{figure}
\epsscale{1.0}
\plotone{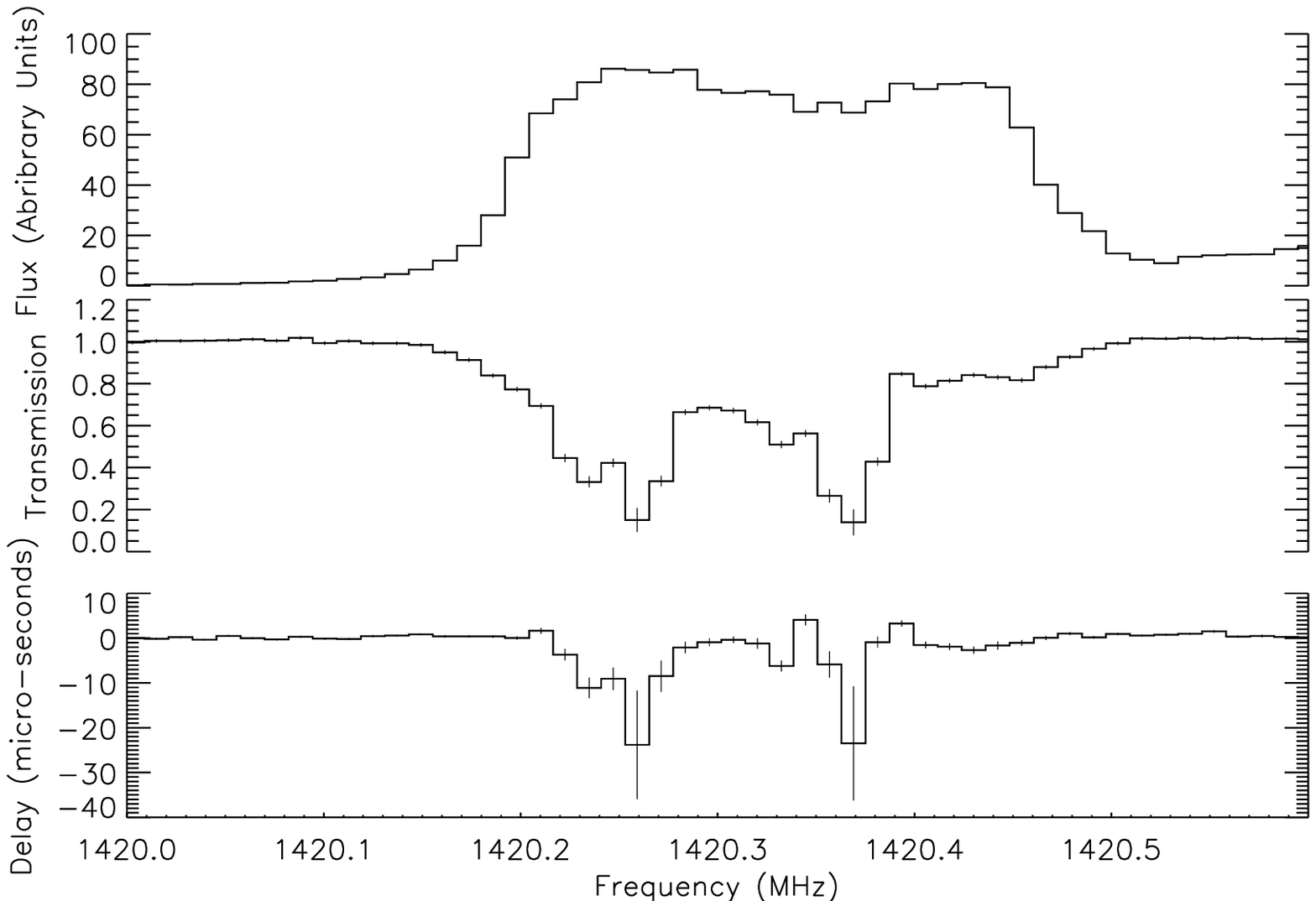}
\caption{\label{fig3}The measured neutral hydrogen emission spectrum
  (top), absorption spectrum (middle), and delay spectrum
  (bottom). The free electron dispersive effects were removed from the
  data by subtracting the best fit linear model to the delay data using
  the off-resonance frequencies. The $1-\sigma$ error bars
  take into account the increased system temperature due to the line
  emission as well as the reduced signal due to the absorption
  spectrum.}
\end{figure}

The absorption spectrum in \ref{fig3} contains both broad and
narrow-line features. There appears to be about four narrow line
features whose peaks line up with the peaks seen in the delay
spectrum. Taking the full-width-half-maximum (FWHM) of these features
to be of order two bins (23 kHz), which corresponds to a kinetic
temperature of about 500 K, the expected delay given by equation
\ref{delay_peak} is of order 10 microseconds. This is consistent with
the observed delay spectrum. Note that the two largest peaks in the
delay spectrum have the largest errors since the absorption is the
greatest at this point. If the delay is actually at 25 microseconds,
as suggest by the data, then there would have to exist narrow-band,
$\approx 10$ kHz, unresolved clouds at these locations. The necessary kinetic
temperatures would have to be about 100 K.

There are also two broad features seen in the data, the first centered
at about 1420.3 MHz, and the other starts just above 1420.4 MHz. It is
not possible to determine the kinetic temperatures of the clouds in
these regions from the absorption spectrum alone since one cannot
measure a line-width. One can use equation \ref{delay_peak} together
with the delay spectrum to estimate the cloud bandwidths and kinetic
temperatures. For the region centered at 1420.3 MHz, the measured
optical depth is .4 and the delay is .9 microseconds. If this region
were made up of a set of clouds with kinetic temperatures of the same
order as the narrow-line features (i.e. 500 K), but with smaller
optical depths,the measured delays would have to be of order 2.5
microseconds, a factor of over 2.5 times the measured value. Using
equation \ref{delay_peak} together with the measured optical depth and
delay, the kinetic temperature of a single cloud in this region would
have to be 5000 K. The FWHM of such a cloud is 70 kHz. Hence, a single
hot cloud could explain the data in this region. Considering the
region above 1420.4 MHz, the optical depth in this region is around .2
while the delay is about 3 microseconds. This corresponds to a cloud
temperature of about 100 K and FWHM of 10 kHz. This width is just
under the frequency resolution of the data. Hence, a single high
temperature cloud in this region cannot explain the data. It must be
made up of about 5 unresolved clouds each with temperatures of about
100 K. It should be noted that this is consistent with high spectral
resolution observations of other pulsars that have revealed the
existence of absorption features as narrow as 2 kHz
\citep{jkw+03,swh+03,fwc+94}.

Given the signal-to-noise ratio and frequency resolution of the
current data, it is difficult to make any definitive statements about
the underlying cloud structure aside from the estimates made
above. Also, the analysis presented here assumes distinct HI clouds in
local thermodynamic equilibrium so that the atomic velocity
distribution is given by a Gaussian distribution and the temperature
refers to the thermal velocity width along the line of sight. Future
observations with longer time integrations and higher frequency
resolution together with more sophisticated analysis techniques that
take into account more complicated, perhaps turbulent, velocity
distributions should enable us to determine or at least place bounds
on the number, average temperature, and column density of the neutral
hydrogen clouds in the line of sight to this pulsar. These techniques
can also be applied to a larger set of pulsars. This will allow us to
gain further insight into the global structure of neutral hydrogen in
the Milky-Way.

%Looking at only the absorption spectrum
%in figure \ref{fig3}, one may conclude that there exist a small
%number, of order 2 to 3, clouds along the line of site each with a
%kinetic temperature of about 500K. If this were the case, the expected
%peak delay, estimated as $-\tau_0/2 \pi^{3/2} f_d$, would be $10 \mu
%s$, far below the actual measured value. This descrepancy can be
%resolved if there exists several unresolved HI clouds with low kinetic
%temperatures. If the peak delay is around 20 microseconds, as
%suggested by the data, the frequency width of these structures would
%have to be of order $4 kHz$, which corresponds to 30 K. 

\acknowledgments 

The lead author wants to thank J. Dawson for discussions which
inspired this work, M. Nolan for his help in obtaining the time at
Arecibo, Phil Perillat and Arun Venkataraman for their invaluable
assistance in setting up the observations, R. Braun, S. Johnston, B.
Koribalski, B. Rickett,  D. Stinebring, and J. Weisberg for their insights and
encouragement during the preparation of this manuscript. The
observations were preformed by students involved in the Arecibo Remote
Command Center (ARCC), a program designed to involve students in
research early in their career. This work was supported by a CAREER grant
from the National Science Foundation (award AST 0545837).

%%%%%%%%%%%%%%%%%%%%%%%%%%%%%%%%%%%%%%%%%%%%%%%%%%%%%%%%%%%%%%%%%%%%%%%%%%

\bibliographystyle{apj}

\begin{thebibliography}{22}
\expandafter\ifx\csname natexlab\endcsname\relax\def\natexlab#1{#1}\fi

\bibitem[{{Abramowitz} \& {Stegun}(1970)}]{as70}
{Abramowitz}, M., \& {Stegun}, I.~A. 1970, {Handbook of mathematical functions
  : with formulas, graphs, and mathematical tables}

\bibitem[{{Armstrong} {et~al.}(1995){Armstrong}, {Rickett}, \&
  {Spangler}}]{ars95}
{Armstrong}, J.~W., {Rickett}, B.~J., \& {Spangler}, S.~R. 1995, \apj, 443, 209

\bibitem[{{Burke} \& {Graham-Smith}(2002)}]{bg02}
{Burke}, B.~F., \& {Graham-Smith}, F. 2002, {An Introduction to Radio
  Astronomy: Second Edition}

\bibitem[{{Condon} \& {Shortley}(1963)}]{cs63}
{Condon}, E.~U., \& {Shortley}, G.~H. 1963, {The theory of atomic spectra}

\bibitem[{{Cordes} {et~al.}(1991){Cordes}, {Weisberg}, {Frail}, {Spangler}, \&
  {Ryan}}]{cwf+91}
{Cordes}, J.~M., {Weisberg}, J.~M., {Frail}, D.~A., {Spangler}, S.~R., \&
  {Ryan}, M. 1991, \nat, 354, 121

\bibitem[{{Dogariu} {et~al.}(2001){Dogariu}, {Kuzmich}, \& {Wang}}]{dzw01}
{Dogariu}, A., {Kuzmich}, A., \& {Wang}, L.~J. 2001, \pra, 63, 053806

\bibitem[{{Frail} {et~al.}(1994){Frail}, {Weisberg}, {Cordes}, \&
  {Mathers}}]{fwc+94}
{Frail}, D.~A., {Weisberg}, J.~M., {Cordes}, J.~M., \& {Mathers}, C. 1994,
  \apj, 436, 144

\bibitem[{{Garrett} \& {McCumber}(1970)}]{gm70}
{Garrett}, C.~G., \& {McCumber}, D.~E. 1970, \pra, 1, 305

\bibitem[{{Han} {et~al.}(2006){Han}, {Manchester}, {Lyne}, {Qiao}, \& {van
  Straten}}]{hml+06}
{Han}, J.~L., {Manchester}, R.~N., {Lyne}, A.~G., {Qiao}, G.~J., \& {van
  Straten}, W. 2006, \apj, 642, 868

\bibitem[{{Heiles} {et~al.}(1983){Heiles}, {Kulkarni}, {Stevens}, {Backer},
  {Davis}, \& {Goss}}]{hks+83}
{Heiles}, C., {Kulkarni}, S.~R., {Stevens}, M.~A., {Backer}, D.~C., {Davis},
  M.~M., \& {Goss}, W.~M. 1983, \apjl, 273, L75

\bibitem[{{Jackson}(1975)}]{jac75}
{Jackson}, J.~D. 1975, {Classical electrodynamics}

\bibitem[{{Johnston} {et~al.}(2001){Johnston}, {Koribalski}, {Weisberg}, \&
  {Wilson}}]{jkw+01}
{Johnston}, S., {Koribalski}, B., {Weisberg}, J.~M., \& {Wilson}, W. 2001,
  \mnras, 322, 715

\bibitem[{{Johnston} {et~al.}(2003){Johnston}, {Koribalski}, {Wilson}, \&
  {Walker}}]{jkw+03}
{Johnston}, S., {Koribalski}, B., {Wilson}, W., \& {Walker}, M. 2003, \mnras,
  341, 941

\bibitem[{{Lorimer} \& {Kramer}(2004)}]{lk04}
{Lorimer}, D.~R., \& {Kramer}, M. 2004, {Handbook of Pulsar Astronomy}

\bibitem[{{Manchester} \& {Taylor}(1977)}]{mt77}
{Manchester}, R.~N., \& {Taylor}, J.~H. 1977, {Pulsars.}

\bibitem[{{Schweinsberg} {et~al.}(2006){Schweinsberg}, {Lepeshkin}, {Bigelow},
  {Boyd}, \& {Jarabo}}]{slb+06}
{Schweinsberg}, A., {Lepeshkin}, N.~N., {Bigelow}, M.~S., {Boyd}, R.~W., \&
  {Jarabo}, S. 2006, Europhysics Letters, 73, 218

\bibitem[{{Sommerfeld}(1954)}]{som54}
{Sommerfeld}, A. 1954, {Optics Lectures on Theortical Physics, Vol. IV}

\bibitem[{{Stanimirovi{\'c}} {et~al.}(2003){Stanimirovi{\'c}}, {Weisberg},
  {Hedden}, {Devine}, \& {Green}}]{swh+03}
{Stanimirovi{\'c}}, S., {Weisberg}, J.~M., {Hedden}, A., {Devine}, K.~E., \&
  {Green}, J.~T. 2003, \apjl, 598, L23

\bibitem[{{Wang} {et~al.}(2000){Wang}, {Kuzmich}, \& {Dogariu}}]{wkd00}
{Wang}, L.~J., {Kuzmich}, A., \& {Dogariu}, A. 2000, \nat, 406, 277

\bibitem[{{Weisberg} {et~al.}(2005){Weisberg}, {Johnston}, {Koribalski}, \&
  {Stanimirovi{\'c}}}]{wjk+05}
{Weisberg}, J.~M., {Johnston}, S., {Koribalski}, B., \& {Stanimirovi{\'c}}, S.
  2005, Science, 309, 106

\bibitem[{{Weisberg} {et~al.}(1980){Weisberg}, {Rankin}, \&
  {Boriakoff}}]{wrb80}
{Weisberg}, J.~M., {Rankin}, J., \& {Boriakoff}, V. 1980, \aap, 88, 84

\bibitem[{{Weisberg} {et~al.}(2008){Weisberg}, {Stanimirovi{\'c}}, {Xilouris},
  {Hedden}, {de la Fuente}, {Anderson}, \& {Jenet}}]{wsx+08}
{Weisberg}, J.~M., {Stanimirovi{\'c}}, S., {Xilouris}, K., {Hedden}, A., {de la
  Fuente}, A., {Anderson}, S.~B., \& {Jenet}, F.~A. 2008, \apj, 674, 286

\end{thebibliography}

\end{document}